\documentclass{osa-article}

\journal{osac}
\articletype{Research Article}
\usepackage{amsmath,bm}
\usepackage{cleveref}
\usepackage{textcomp}
\usepackage{placeins}

\newcommand{\beq}{\begin{equation}}
\newcommand{\eeq}{\end{equation}}

\newcommand{\eps}{\varepsilon}
\begin{document}

\title{Direct retrieval method of the effective permittivity and permeability of bulk semi-infinite metamaterials by variable-angle spectroscopic ellipsometry}

\author{Quentin Flamant,\authormark{1,2} Daniel Torrent,\authormark{1,2,4} Sergio Gomez-Gra\~{n}a,\authormark{3,5} Alexander N. Grigorenko,\authormark{6} Vasyl G. Kravets,\authormark{6} Philippe Barois,\authormark{1,2} Virginie Ponsinet,\authormark{1,2} and Alexandre Baron\authormark{1,2,*}}

\address{
\authormark{1}Universit\'e de Bordeaux, Centre de Recherche Paul Pascal (CRPP), UMR 5031, F-33600 Pessac, France\\
\authormark{2}CNRS, Centre de Recherche Paul Pascal (CRPP), UMR 5031, F-33600 Pessac, France\\
\authormark{3}Universit\'e de Bordeaux, Institut de Chimie de la Mati\`ere Condens\'ee de Bordeaux  (ICMCB), UMR 5026, F-33600 Pessac, France\\
\authormark{4}Currently at GROC, Institut de Noves Tecnologies de la Imatge (INIT), Universitat Jaume I, Castell\'on 12071, Spain\\
\authormark{5}Currently at Departamento de Qu\'{i}mica F\'{i}sica, CINBIO, Universidade de Vigo, 36310 Vigo, Spain
\authormark{6}University of Manchester, School of Physics \& Astronomy, Manchester, M13 9PL, United Kingdom
}

\email{\authormark{*}baron@crpp-bordeaux.cnrs.fr} %% email address is required

% \homepage{http:...} %% author's URL, if desired

%%%%%%%%%%%%%%%%%%% abstract %%%%%%%%%%%%%%%%
%% [use \begin{abstract*}...\end{abstract*} if exempt from copyright]

\begin{abstract}
In this work, we present a simple method for the direct retrieval of the effective permittivity and permeability of a bulk semi-infinite metamaterial from variable-angle spectroscopic ellipsometry measurements. Starting from the well-known Fresnel equations, we derive an analytical expression in which unknown coefficients are fitted to the experimental data using a linear regression model. The effective permittivity and permeability are then determined by solving a simple system and the correct solution is selected based on physical criteria. As an example, the method is applied to the case of a self-assembled metamaterial exhibiting strong isotropic optical magnetism.
\end{abstract}

%%%%%%%%%%%%%%%%%%%%%%%%%%  body  %%%%%%%%%%%%%%%%%%%%%%%%%%
\section{Introduction}
Spectroscopic ellipsometry measures polarized light reflected from the surface of a material or thin-film and lets two polarization directions interfere to produce a measurable signal as a function of wavelength. The measured quantities enable the fitting or retrieval of the optical and sometimes structural properties of the material. Variable-angle spectroscopic ellipsometry is the application of this method to different incidence angles. It ultimately enables a full characterization of the optical index $n=n'+in"$. The measured quantity is typically the ellipsometric ratio
\begin{equation}
    \rho = \frac{r_p}{r_s} = \tan\psi e^{-i\Delta}
    \label{eq:rho}
\end{equation}
where $r_p$ ($r_s$) is the p-polarized (s-polarized) reflection coefficient and $(\psi,\Delta)$ are the ellipsometric angles. Note that the sign within the exponential function depends on the convention used for the definition of the phase of light waves. Here we assume a time dependence of the form $\exp(-i\omega t)$. When light is reflected from a bulk semi-infinite homogeneous material, the reflection coefficients are simply given by the Fresnel equations and the optical index can be retrieved directly from $\rho$\cite{fujiwara2007}
\begin{equation}
    n = n_{\mathrm{background}}\sin\theta_1\left[1+\left(\frac{1-\rho}{1+\rho}\right)^2\tan^2\theta_1\right]^{1/2}
    \label{eq:N}
\end{equation}
where $\theta_1$ is the angle of incidence and $n_{\mathrm{background}}$ is the index of the background (usually it is air, so $n_{\mathrm{background}} \approx 1$). Since ellipsometry measures two values -- namely the real and imaginary parts of $\rho$ -- the retrieval procedure unambiguously determines both the real and imaginary parts of $n$.

For a little over fifteen years now, artificially structured materials composed of subwavelength inclusions arranged into two- and three-dimensional ensembles have been engineered with unusual and sometimes fascinating properties. When they can be described with homogenized effective electromagnetic parameters, such materials are usually referred to as \textit{metamaterials}\cite{smith2000composite, smith2002determination,soukoulis2011past,baron2016invited}. For an interesting perspective on their development, we refer the interested reader to Tretyakov's recent article \cite{tretyakov2016personal}. 

The unusual properties of metamaterials make their characterization challenging, since the basic assumptions of classical models are often broken. Let us consider the case of a nanoparticle composite, which is the simplest type of metamaterial. In cases where the particles exhibit a dipolar plasmon resonance, which is purely electric, the electromagnetic response of the ensemble is fully characterized by an effective electric permittivity $\varepsilon=n^2$. If the metamaterial is thick and absorbing enough to be considered as semi-infinite, eq. \ref{eq:N} can then be used to determine the optical index \cite{baron2013bulk}. However, this simple approach does not hold if the particles exhibit a magnetic dipolar response. As some of the authors of the present paper demonstrated, it is indeed possible to achieve magnetism at optical frequencies in a bottom-up self-assembled metamaterial consisting of a three-dimensional homogeneous ensemble of magnetic nanoparticles \cite{gomez2016hierarchical}. As a result, the metamaterial is described not only by an effective electric permittivity $\varepsilon$ but also by an effective magnetic permeability $\mu$.

The classical approach to retrieve the optical parameters of metamaterials is the \textit{S-parameters retrieval method} inherited from microwave research. It consists in measuring the normal incidence reflection and transmission coefficients $r$ and $t$ for the electric field (the so-called \textit{scattering parameters}) in order to calculate the effective electromagnetic constants using simple algebraic equations assuming that the thickness of the sample is known \cite{smith2002determination, Smith2005}. This elegant, simple and powerful method is extensively used by the metamaterial community. However, it has several major drawbacks. First, in the case of visible optics, the accurate measurement of the phase of transmitted light can be problematic for thick samples and is usually complicated by the presence of a substrate. Second, an unreserved application of this method to thin layers of meta-atoms is not correct as the ``geometrical'' thickness of a metamaterial sample could be different from its ``optical'' thickness due to the near-fields associated with the meta-atoms. Third, it implicitly assumes that the metamaterial under study can be described by the effective medium theory and hence that it can be replaced by an effective optical layer, whose reflection and transmission coefficients can be computed from Fresnel equations. This is a strong assumption and for many metamaterials it requires confirmation. For example, the \textit{S-parameters retrieval method} leads to a negative refractive index for a sandwich structure consisting of thin layers of 50 nm glass and 10 nm gold. This spurious result comes from the fact that such a sandwich cannot be described by the effective medium theory (for details see \cite{kravets2010}). Indeed, when the effective medium theory cannot be applied (\textit{e.g}., in the presence of spatial dispersion effects), the retrieved optical parameters are angle-dependent wave parameters rather than global material parameters \cite{Menzel2008}.

In our previous work, we proposed a retrieval method based on variable-angle spectroscopic ellipsometry measurements that overcomes the drawbacks of the celebrated \textit{S-parameters retrieval method} in the case of an optically thick metamaterial. Indeed, by measuring the electric field reflection coefficients at different angles and using simple algebra we not only can retrieve the optical constants, but also check the applicability of the effective field theory to the studied metamaterial (\textit{i.e.}, confirm these constants). To our knowledge this was the first time that such a method was proposed and applied experimentally to measure and confirm $\varepsilon$ and $\mu$ in optics. The purpose of this paper is therefore to provide a detailed description of the retrieval procedure complemented with a tutorial case using the data previously published by Gomez-Gra\~{n}a et al. \cite{gomez2016hierarchical}. 

\section{Retrieval method}
\FloatBarrier
In a variable angle spectroscopic ellipsometry experiment, the material parameters need to be determined through a retrieval procedure, which requires a description of how the reflection (or the transmission) coefficients behave depending on the polarization and the angle of incidence.

\begin{figure}
\centering
\includegraphics[width=\columnwidth]{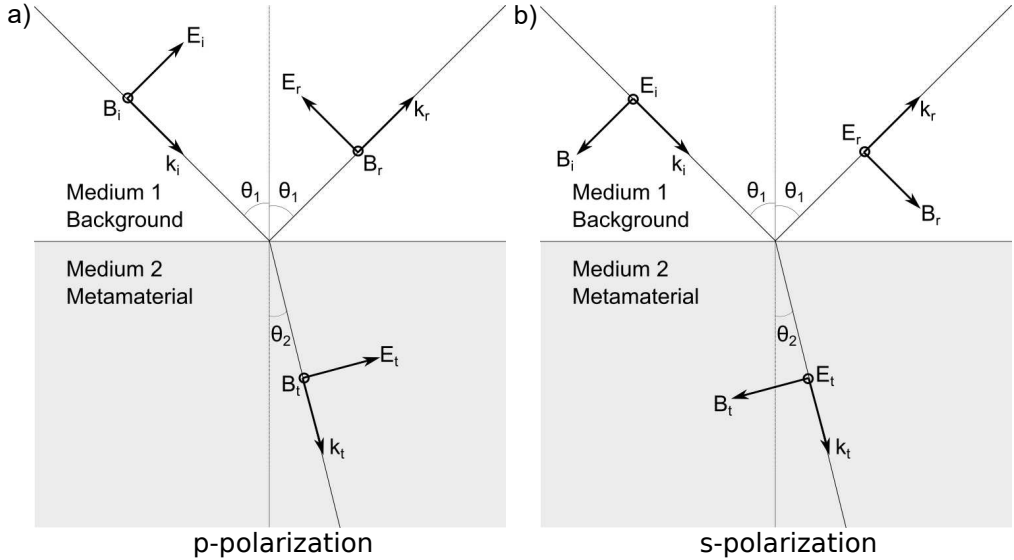}
\caption{\textnormal{Electric field $E$, magnetic induction $B$ and wavevector $k$ corresponding to the incident, reflected and transmitted wave for a) $p-$polarization and b) $s-$polarization. $B$ in a) and $E$ in b) are perpendicular to the plane of the paper and are pointing to the reader. NB: the case represented here is that for which $\theta_2$ is positive, which corresponds to a positive refractive index, $n_2$. However the retrieval method presented in this paper is equally valid for a negative refractive index.}}
\label{fig_schema}
\end{figure}

Let us assume that a monochromatic electromagnetic plane wave with a time dependence of the form $\exp(-i\omega t)$ is propagating through an isotropic medium 1 with a permittivity $\eps_1$ and a permeability $\mu_1$ and is incident on a planar interface with an isotropic medium 2 with a permittivity $\eps_2$ and a permeability $\mu_2$ (see Fig. \ref{fig_schema}) at an angle of incidence $\theta_1$. Medium 1 is the background (\textit{e.g}., air), medium 2 is the metamaterial with unknown electromagnetic properties and $\theta_2$ is the refraction angle inside the metamaterial.

In the present work, we consider an optically thick passive metamaterial slab, which we model as a semi-infinite medium. Therefore, the reflection coefficient for the $p-$polarization and the $s-$polarization can be obtained from the well-known Fresnel equations \cite{hecht2017optics}:
\begin{eqnarray}
\label{eq:r_p}
r_p=\frac{Z_1\cos{\theta_1}-Z_2\cos{\theta_2}}{Z_1\cos{\theta_1}+Z_2\cos{\theta_2}}\\
\label{eq:r_s}
r_s=\frac{Z_2\cos{\theta_1}-Z_1\cos{\theta_2}}{Z_2\cos{\theta_1}+Z_1\cos{\theta_2}}
\end{eqnarray}
where $Z_1=\sqrt{\mu_1 / \eps_1}$ and $Z_2=\sqrt{\mu_2 / \eps_2}$  are the impedances in media 1 and 2, respectively.
Combining eq. \ref{eq:rho}, \ref{eq:r_p} and \ref{eq:r_s} gives
\beq
\label{eq:rho_2}
\rho=\frac{(Z_1\cos{\theta_1}-Z_2\cos{\theta_2})(Z_2\cos{\theta_1}+Z_1\cos{\theta_2})}
{(Z_1\cos{\theta_1}+Z_2\cos{\theta_2})(Z_2\cos{\theta_1}-Z_1\cos{\theta_2})}
\eeq
which after some algebra leads to
\beq
\frac{1-\rho}{1+\rho}=\frac{\cos{\theta_1}\cos{\theta_2}}{\cos^2{\theta_1}-\cos^2{\theta_2}}
\frac{Z_2^2-Z_1^2}{Z_1 Z_2}
\eeq
Using Snell relation $n_1\sin{\theta_1}=n_2\sin{\theta_2}$ and performing some additional calculation, this expression is transformed into
\begin{equation}
\label{eq:rho_exp}
\left(\frac{1-\rho}{1+\rho}\right)^2 \frac{\sin^4{\theta_1}}{\cos^2{\theta_1}} 
= \left(1-\frac{n_1^2}{n_2^2}\sin^2{\theta_1}\right)
\left(\frac{n_2^2}{n_1^2-n_2^2}\right)^2
\left(\frac{Z_2^2-Z_1^2}{Z_1 Z_2}\right)^2
\end{equation}
Using the relations $n_i=\sqrt{\mu_i \eps_i}$ and $Z_i = \sqrt{\mu_i / \eps_i}$ in eq. \ref{eq:rho_exp} yields
\begin{equation}
\label{eq:rho_exp2}
\left(\frac{1-\rho}{1+\rho}\right)^2 \frac{\sin^4{\theta_1}}{\cos^2{\theta_1}}
=
\left(\frac{\eps_2 \mu_1 - \eps_1 \mu_2}{\eps_1 \mu_1 - \eps_2 \mu_2}\right)^2
\left(\frac{\eps_2 \mu_2}{\eps_1 \mu_1} - \sin^2{\theta_1} \right)
\end{equation}

Defining $\eps=\eps_2 / \eps_1$ and $\mu=\mu_2 / \mu_1$ as the permittivity and permeability of the metamaterial relative to those of the background, this expression simplifies to
\begin{equation}
\label{eq:rho_exp3}
\left(\frac{1-\rho}{1+\rho}\right)^2 \frac{\sin^4{\theta_1}}{\cos^2{\theta_1}} =
\left(\frac{\eps - \mu}{1 - \eps \mu}\right)^2
\left(\eps \mu - \sin^2{\theta_1}\right)
\end{equation}

In this equation, it is implicitly assumed that $\rho \neq -1$ and $\eps \neq 1/\mu$, which is verified provided that $Z_1 \neq 0$, $Z_2 \neq 0$, $n_1 \neq n_2$ and $n_1 \neq -n_2$. If medium 1 is vacuum or air, $\eps_1 \approx 1$ and $\mu_1 \approx 1$, so $\eps=\eps_2$ and $\mu=\mu_2$. Since this is usually the case in an ellipsometry experiment, we shall assume it to be true in the rest of the analysis. Nevertheless, if $\eps_1$ and $\mu_1$ differ from unity but are known and real-valued (\textit{i.e.}, medium 1 is lossless), it is still possible to extract $\eps_2$ and $\mu_2$ using the same method.

Equation \ref{eq:rho_exp3} can be cast in the form of a linear regression model of the form
\beq
\label{eq:eq_Y_AX_B}
Y=AX+B
\eeq
where the experimental variables are
\begin{align}
\label{eq:eq_X}
X&=-\sin^2\theta_1 \\
\label{eq:eq_Y}
Y&=\left(\frac{1-\rho}{1+\rho}\right)^2\frac{\sin^4\theta_1}{\cos^2\theta_1}
\end{align}
and the parameters to be determined are
\begin{eqnarray}
\label{eq:AB1}
A&=\left(\frac{\eps-\mu}{1-\eps\mu}\right)^2 \\
\label{eq:AB2}
B&=\left(\frac{\eps-\mu}{1-\eps\mu}\right)^2\eps\mu
\end{eqnarray}
The metamaterial is usually lossy while the background is lossless, therefore $\rho$, $\eps$ and $\mu$ are complex quantities, and $\theta_1$ is real. Consequently, $Y$, $A$ and $B$ are complex and $X$ is real. The linear regression model can thus be split into two equations:
\begin{align}
\label{eq:real_part_eq}
\mathit{Re(Y)}&=\mathit{Re(A)}X+\mathit{Re(B)} \\
\label{eq:im_part_eq}
\mathit{Im(Y)}&=\mathit{Im(A)}X+\mathit{Im(B)}
\end{align}
For a given wavelength, the set of measurements $\rho=\rho(\theta_1)$ provides the data to fit the model, from which we can retrieve the properties of the material. It must be pointed out that, without performing any inversion, for $\eps \neq\ \mu$ the square of the refractive index can be computed from
\beq
\label{eq:n2}
\eps\mu=n^2=\frac{A}{B}
\eeq
which provides a robust measurement of this quantity. Note that the case $\eps = \mu$ yields $A = B = 0$ but is trivial, since it provides an additional relation and the experimental data can be fitted directly to Snell's law. Finally, both $\eps$ and $\mu$ can be obtained as a function of A and B after inversion of equations \eqref{eq:AB1} and \eqref{eq:AB2}. Some algebra shows that $\eps$ can be determined from the solution of the following equation
\beq
\label{eq:epsilon}
\eps^2\pm\eps\sqrt{A}(1-B/A)-B/A=0
\eeq
The $\pm$ sign comes from the inversion of equation \eqref{eq:AB1}, so that there are actually two of the above equations, each of which provides two solutions for $\eps$, which yields to a total of four solutions
\begin{align}
\label{eq:eps_a_plus}
\eps_a^+ &= \frac{-\sqrt{A}(1-\frac{B}{A})+\sqrt{\Delta}}{2} \\
\label{eq:eps_b_plus}
\eps_b^+ &= \frac{-\sqrt{A}(1-\frac{B}{A})-\sqrt{\Delta}}{2} \\
\label{eq:eps_a_minus}
\eps_a^- &= \frac{\sqrt{A}(1-\frac{B}{A})+\sqrt{\Delta}}{2} \\
\label{eq:eps_b_minus}
\eps_b^- &= \frac{\sqrt{A}(1-\frac{B}{A})-\sqrt{\Delta}}{2}
\end{align}
where $\eps_{a,b}^+$ are the solutions for eq. \eqref{eq:epsilon} with the + sign, $\eps_{a,b}^-$ are the solutions for eq. \eqref{eq:epsilon} with the - sign and
\beq
\label{eq:refractive_index}
\Delta=\frac{(A-B)^2+4B}{A}
\eeq

From equation \eqref{eq:n2}, we obtain the value of $\mu$ corresponding to each $\eps$. The only remaining problem is the determination of which of the four couples of solutions ($\eps$, $\mu$) is the physical one. A possible approach is to compute the corresponding refractive index, taking care of choosing the correct sign for the square root $\sqrt{\mu \eps}$. This can be done easily using the following formula based on causality \cite{Grigorenko2006}:
\beq
n=\sqrt{|\eps||\mu|}\exp \left[i\left(\frac{\delta_\epsilon+\delta_\mu}{2}\right)\right]
\mathrm{sgn}\left[\cos {\left(\frac{\delta_\epsilon-\delta_\mu}{2}\right)}\right]
\eeq
where $\delta_\epsilon$ and $\delta_\mu$ denote the arguments of the complex quantities $\eps$ and $\mu$ and sgn denote the sign. 

For a given selection of the sign of $\sqrt{A}$ in eq. \eqref{eq:epsilon}, it must be pointed out that the product $\eps_a^+ \eps_b^+$ (or $\eps_a^- \eps_b^-$) of the two solutions is $-A/B$. Moreover, according to equation \eqref{eq:n2}, the product of $\eps$ and $\mu$ is $A/B$. Therefore, if one solution (\textit{e.g.}, $\eps_a^+$) gives $\eps$, the other one (\textit{e.g.}, $\eps_b^+$) gives $-\mu$. Some simple calculation shows then that the two possible solutions for $\eps$ lead to an opposite sign of $\cos {\left(\frac{\delta_\epsilon-\delta_\mu}{2}\right)}$, and thus to an opposite sign of the refractive index.

Since the medium considered is passive, the imaginary part of the refractive index must be positive (for the time dependence of the form $\exp(-i\omega t)$ considered here). Two couples of solution are thus eliminated, one for each choice of the sign of $\sqrt{A}$. For a passive medium the real part of the impedance must also be positive, therefore an additional verification can be done by computing $Z=n / \eps$. 

Finally, because we retrieve the parameters as a function of frequency, it is easy to determine which of the two remaining solutions is the correct one, using the fact that $\mu$ must be nearly 1 far away from the resonance region. Besides, the real part of both $\eps$ and $\mu$ must be positive far from the resonance, which can also be checked. Note that since the magnetic response of a metamaterial is the result of spatial dispersion in a non-magnetic material, there is no theoretical restriction on the sign of the imaginary parts of $\eps$ and $\mu$. In practice, the imaginary part of $\eps$ is expected to be positive but the imaginary part of $\mu$ may be negative provided that the overall electromagnetic energy dissipated in the material is positive.
\FloatBarrier

%%%%%%%%%%%%%%%%%%%%%%%%%%%%%%%%%%%%%%%%%%%%%%%%%%%%%%%%%%%%%%%%%%%%%%%%%%%%%%%%%%%%

\section{Tutorial case}
\FloatBarrier
As an example of application for the retrieval method, let us use our previously published data on a self-assembled metamaterial consisting of raspberry-like magnetic nanoclusters exhibiting strong isotropic optical magnetism in visible light \cite{gomez2016hierarchical}.

\begin{figure}
\centering
\includegraphics[width=\columnwidth]{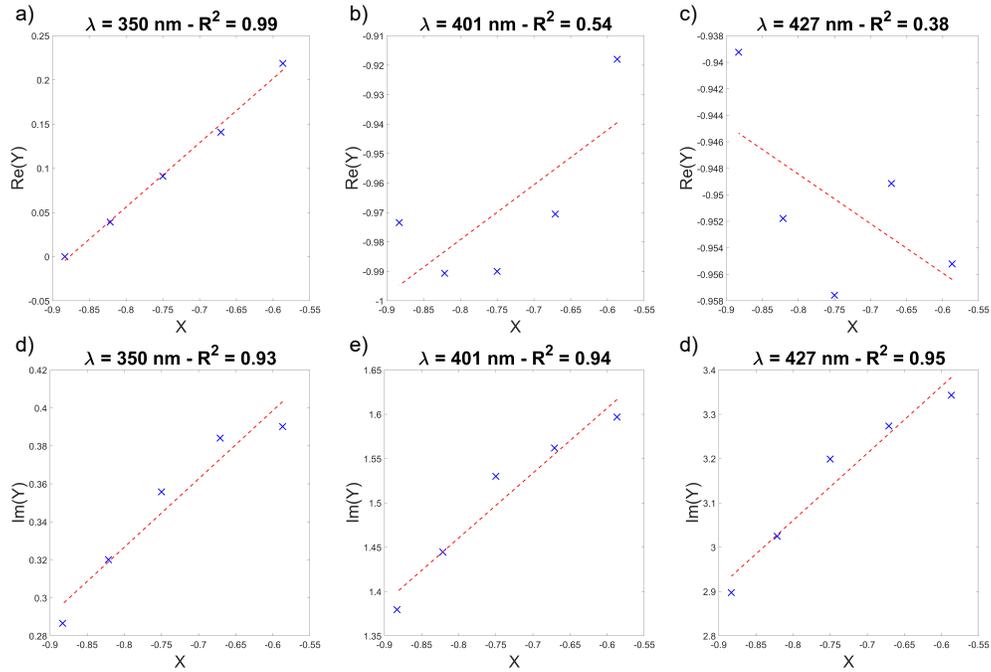}
\caption{\textnormal{Examples showing the line of best fit (red line) for experimental data from \cite{gomez2016hierarchical} (blue crosses) based on the linear regression model described in \crefrange{eq:eq_Y_AX_B}{eq:im_part_eq} for three different wavelengths: a), d) $\lambda = 350 \, \mathrm{nm}$; b), e) $\lambda = 401\, \mathrm{nm}$ ; c), f) $\lambda = 427 \, \mathrm{nm}$. The experimental data was collected at five angles of incidence $\theta_1$ (50, 55, 60, 65 and 70 \textdegree). Graphs a), b), c) represent the real part of the experimental variable Y, while graphs d), e) and f) represent the imaginary part. The linear model fits well Re(Y) at $\lambda = 350 \, \mathrm{nm}$ but poorly at $\lambda = 401 \, \mathrm{nm}$ and $\lambda = 427 \, \mathrm{nm}$. The linear model fits reasonably well Im(Y) in the three cases.}}
\label{fig_fit}
\end{figure}

\begin{figure}
\centering
\includegraphics[width=\columnwidth]{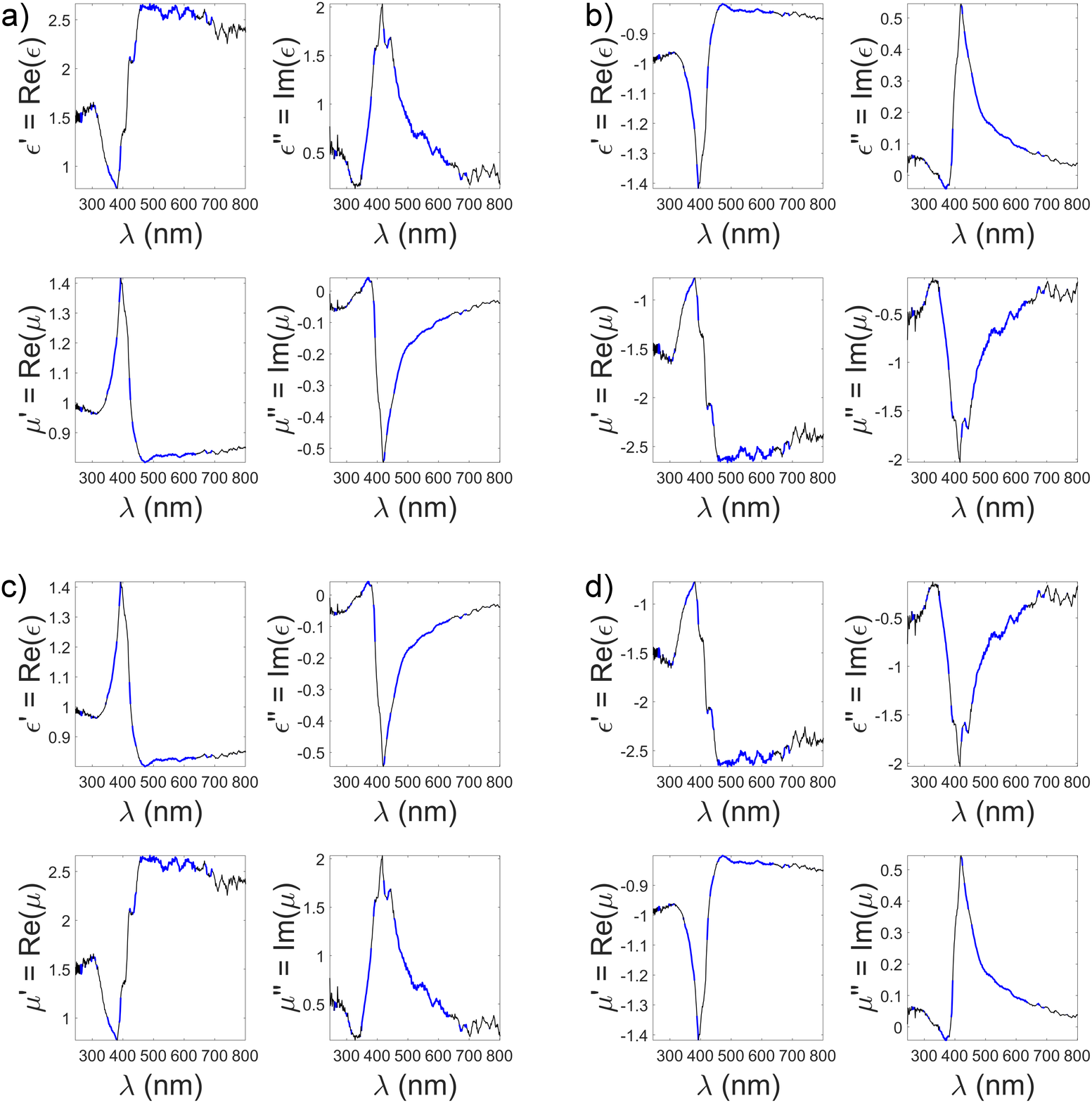}
\caption{\textnormal{Four possible solutions for the values of the real and imaginary parts of the permittivity $\eps$ and the permeability $\mu$ obtained using the retrieval method presented in this manuscript on ellipsometric data from \cite{gomez2016hierarchical}. The graphs correspond respectively to the solution: a) $\eps_a^+$ (eq. \eqref{eq:eps_a_plus}), b) $\eps_b^+$ (eq. \eqref{eq:eps_b_plus}), c) $\eps_a^-$ (eq. \eqref{eq:eps_a_minus}) and d) $\eps_b^-$} (eq. \eqref{eq:eps_b_minus}). The thick blue sections of the curves correspond to wavelengths for which both coefficients of determination $R^2$ of the linear regressions (eq. \eqref{eq:real_part_eq} and eq. \eqref{eq:im_part_eq}) are higher than 0.9, indicating that the model provides a satisfactory description of the data.}
\label{fig_eps_mu}
\end{figure}

\begin{figure}
\centering
\includegraphics[width=\columnwidth]{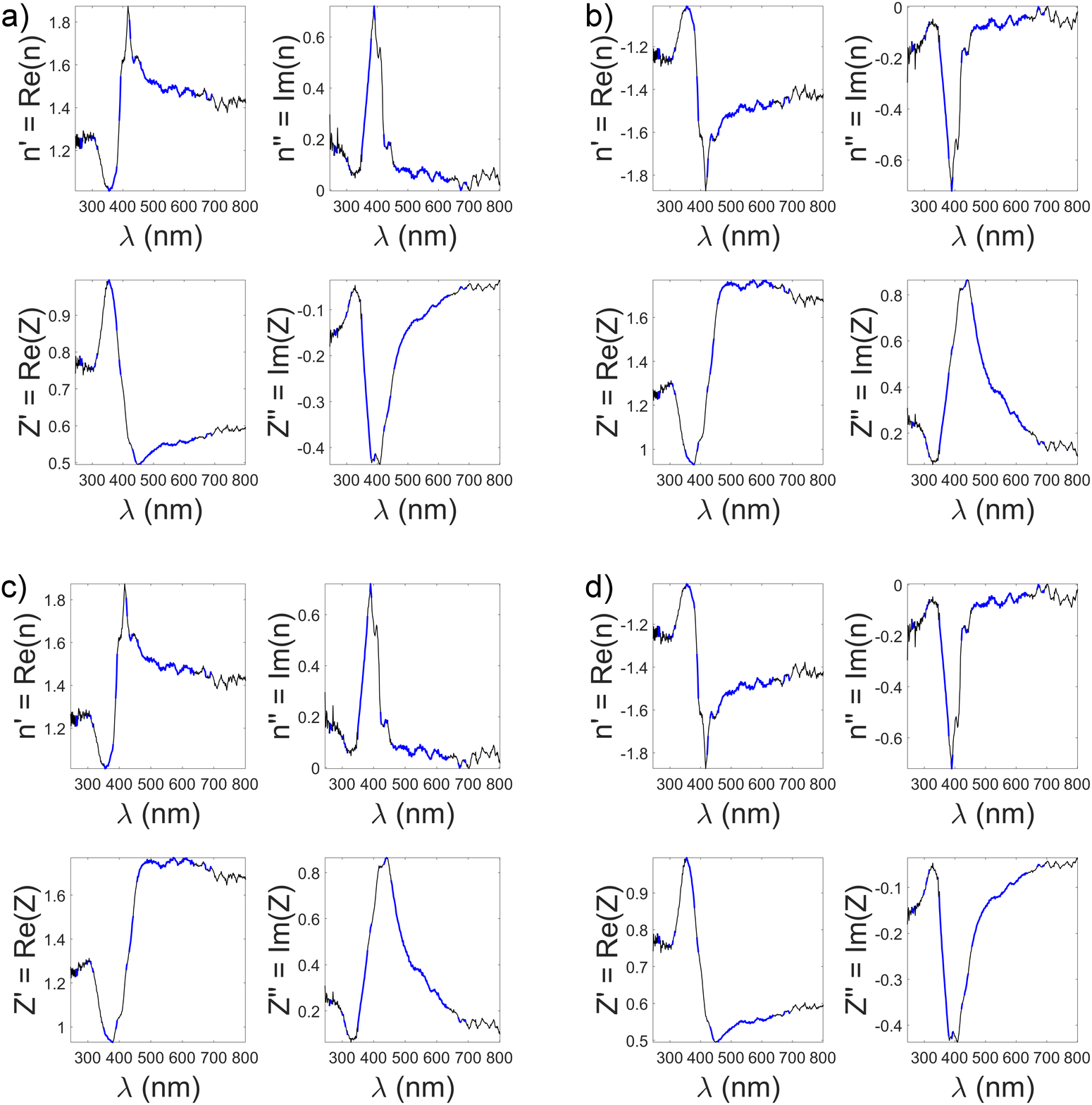}
\caption{\textnormal{Four possible solutions for the values of the real and imaginary parts of the refractive index $n$ and the impedance $Z$ obtained using the retrieval method presented in this manuscript on ellipsometric data from \cite{gomez2016hierarchical}. The graphs correspond respectively to the solution: a) $\eps_a^+$ (eq. \eqref{eq:eps_a_plus}), b) $\eps_b^+$ (eq. \eqref{eq:eps_b_plus}), c) $\eps_a^-$ (eq. \eqref{eq:eps_a_minus}) and d) $\eps_b^-$} (eq. \eqref{eq:eps_b_minus}). The thick blue sections of the curves correspond to wavelengths for which both coefficients of determination $R^2$ of the linear regressions (eq. \eqref{eq:real_part_eq} and eq. \eqref{eq:im_part_eq}) are higher than 0.9, indicating that the model provides a satisfactory description of the data.}
\label{fig_n_Z}
\end{figure}

For each wavelength, we get experimental complex values of $\rho$ as a function of the angle of incidence $\theta_1$. From eq. \ref{eq:eq_X} and eq. \ref{eq:eq_Y}, we then determine the experimental variables $X$ and $Y$. Note that both variables depend on $\theta_1$. Subsequently, we perform a linear regression on the real and the imaginary part of $Y$ (eq. \ref{eq:real_part_eq} and \ref{eq:im_part_eq}), which gives us a value for the parameters $A$ and $B$ at a given wavelength. An example of successful regression is shown in Figure \ref{fig_fit}-a),d). From \crefrange{eq:eps_a_plus}{eq:eps_b_minus}, we then compute the four possible solutions for the permittivity $\eps$ and from eq. \eqref{eq:n2} the corresponding permeability $\mu$ values, as shown in Figure \ref{fig_eps_mu}. Finally, we determine the refractive index $n$ from eq. \ref{eq:refractive_index}, and deduce the impedance $Z$, as shown in Figure \ref{fig_n_Z}.

The next task is to select the correct solution. We see that $\eps_b^+$ (Figure \ref{fig_n_Z}-b)) and $\eps_b^-$ (Figure \ref{fig_n_Z}-d)) lead to unphysical negative values of the imaginary part of the refractive index and of the real part of the impedance, which eliminates two possibilities. We then remark that $\eps_a^-$ does not lead to a $\mu$ value that gets close to 1 far from the resonance (Figure \ref{fig_eps_mu}-c)). The correct solution is thus $\eps_a^+$ (Figure \ref{fig_eps_mu}-a), Figure \ref{fig_n_Z}-a)).

\begin{figure}
\centering
\includegraphics[width=\columnwidth]{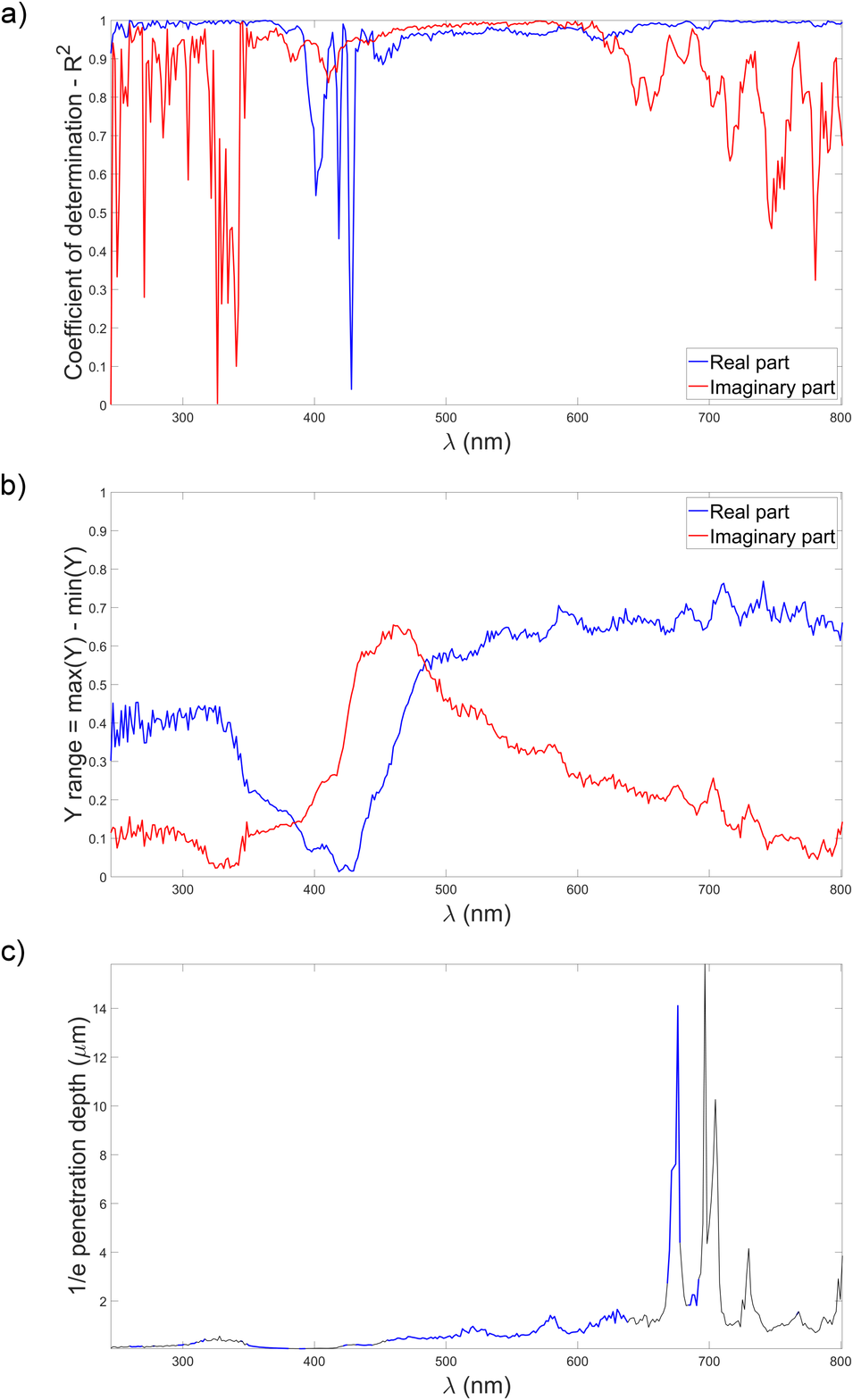}
\caption{\textnormal{a) Coefficient of determination for the linear regression on the real part (eq. \eqref{eq:real_part_eq}) and the imaginary part (eq. \eqref{eq:im_part_eq}) of Y as a function of wavelength, indicating the goodness of the fit for the experimental data from \cite{gomez2016hierarchical}. b) Experimental range for the real and the imaginary part of the variable Y as a function of wavelength. c) 1/e penetration depth calculated as $\delta_p = \lambda / 4 \pi n"$ as a function of wavelength. The thick blue sections of the curve corresponds to wavelengths for which both coefficients of determination $R^2$ of the linear regressions (eq. \eqref{eq:real_part_eq} and eq. \eqref{eq:im_part_eq}) are higher than 0.9, indicating that the model provides a satisfactory description of the data.}}
\label{fig_goodness_of_fit}
\end{figure}

The last remaining question is whether the model correctly describes the experimental data. To address it, we compute the coefficient of determination, which allows us to assess the goodness of the linear fit (Figure \ref{fig_goodness_of_fit}-a)). It appears that the fit of the real part of Y is overall good ($R^2 > 0.9$) except for three narrow regions: $\approx$ 393 nm - 410 nm (1), $\approx$ 415 nm - 420 nm (2) and $\approx$ 425 nm - 431 nm (3). Region (1) lies in the wavelength range where the system exhibits a strong resonance and the maximum $\mu$ value. The artificial magnetic response of metamaterials is typically interpreted as a ``non-local'' effect, which means that it is due to the explicit dependence of the effective dielectric constant on the wavenumber $k$. This dependence on $k$ is not taken into account in our model and goes out of the scope of this article, but it explains why the linear regression fails to correctly fit the data. Indeed, looking for instance at the case $\lambda = 401\, \mathrm{nm}$, we see that the distribution of the experimental data points suggests a quadratic rather than linear relation (Figure \ref{fig_fit}-b). On the other hand, the poor fit in region (2) and (3) is simply due to the fact that at these wavelengths the experimental range of variation of Re(Y) is very small (Figure \ref{fig_goodness_of_fit}-b), which hampers the correct determination of the slope because of the experimental uncertainty. This is substantiated by the erratic distribution of the experimental points, as can be seen at $\lambda = 427\, \mathrm{nm}$ (Figure \ref{fig_fit}-c).

The fit of the imaginary part is good ($R^2 > 0.9$) on a more restricted range (mainly $\approx$ 345 nm - 635 nm with a slight drop between 380 nm and 420 nm). The deviation from the linear model at short wavelengths ($<$ 320 nm) can be attributed to a stronger experimental noise, which explains the erratic behavior observed on $R^2$. Similarly to what happens with Re(Y) in region (2) and (3), the strong drop in $R^2$ observed from 320 nm to 345 nm is merely due to the fact that at these wavelengths the experimental range of variation of Im(Y) becomes small compared to the experimental uncertainty (Figure \ref{fig_goodness_of_fit}-b). The slight drop from 380 nm to 420 nm can be related to the non-local effects close to the resonance that were discussed above. Beyond 635 nm, the poor fits can be explained by the reduced losses far from resonance that cause the penetration depth to increase (Figure \ref{fig_goodness_of_fit}-c): indeed the metamaterial slab is $\approx$ 5 $\mu m$ thick and the semi-infinite medium hypothesis breaks down if absorption is too low. 

\section*{Conclusion}
The method presented in this paper allows a simple and direct retrieval of the effective permittivity and permeability of a bulk semi-infinite isotropic metamaterial from variable-angle spectroscopic ellipsometry measurements. The main restriction on its application is the existence of non-local effects in regions of strong resonance, which are not taken into account by the model and will be the subject of a future work.

\FloatBarrier

\section*{Acknowledgments}
The authors acknowledge support from the LabEx AMADEus (ANR-10-LABX-42) in the framework of IdEx Bordeaux (ANR-10-IDEX-03-02), France. D.T. acknowledges financial support through the ``Ram\'on y Cajal'' fellowship under grant number RYC-2016-21188. S.G.G. acknowledges financial support through the ``Juan de la Cierva - Incorporaci\'{o}n'' fellowship under grant number IJIC-2016-29108. All co-authors thank M. Tr\'eguer-Delapierre and E. Duguet for their valuable input.

%%%%%%%%%% If using BibTeX:
\bibliographystyle{osajnl}
\bibliography{sample}

%%%%%%%%%% If preparing manually:
% \begin{thebibliography}{1}
% \newcommand{\enquote}[1]{``#1''}

% \bibitem{Zhang:14}
% Y.~Zhang, S.~Qiao, L.~Sun, Q.~W. Shi, W.~Huang, L.~Li, and Z.~Yang,
%   \enquote{Photoinduced active terahertz metamaterials with nanostructured
%   vanadium dioxide film deposited by sol-gel method,}
%   {\protect\JournalTitle{Optics Express}} \textbf{22}, 11070--11078 (2014).

% \bibitem{OSA}
% {Optical Society}, \enquote{{OSA Publishing},}
%   \url{http://www.osapublishing.org}.

% \bibitem{FORSTER2007}
% P.~Forster, V.~Ramaswamy, P.~Artaxo, T.~Bernsten, R.~Betts, D.~Fahey,
%   J.~Haywood, J.~Lean, D.~Lowe, G.~Myhre, J.~Nganga, R.~Prinn, G.~Raga,
%   M.~Schulz, and R.~V. Dorland, \enquote{Changes in atmospheric consituents and
%   in radiative forcing,} in \enquote{Climate Change 2007: The Physical Science
%   Basis. Contribution of Working Group 1 to the Fourth assesment report of
%   Intergovernmental Panel on Climate Change,}  S.~Solomon, D.~Qin, M.~Manning,
%   Z.~Chen, M.~Marquis, K.~B. Averyt, M.~Tignor, and H.~L. Miler, eds.
%   (Cambridge University Press, 2007).

% \end{thebibliography}

\end{document}